\documentstyle[prd,aps]{revtex}
\newcommand{\beq}{\begin{equation}}
\newcommand{\beqn}{\begin{eqnarray}}
\newcommand{\eeq}{\end{equation}}
\newcommand{\eeqn}{\end{eqnarray}}
\begin{document}
\draft
\twocolumn[\hsize\textwidth\columnwidth\hsize\csname
@twocolumnfalse\endcsname

\preprint{HUTP-96/A032; astro-ph/9608025}
\title{Supercurvature Modes from Preheating in an Open Universe}
\author{David I. Kaiser}
\address{Lyman Laboratory of Physics, Harvard University, Cambridge, MA
02138 USA}
\date{5 August 1996}
\maketitle
\begin{abstract}
Preheating following the inflationary phase in models of open inflation is
considered.  The most significant difference from a preheating scenario in
flat space is that supercurvature modes can be populated and amplified in
the open universe.  In certain models, such modes can dominate the usual 
resonant particle production, thus altering the ensuing thermal 
history within an open universe.  The accuracy to which the masses and
couplings need to be tuned to produce such supercurvature modes, however,
makes any large deviation from the flat space cases unlikely.
\end{abstract}
\pacs{PACS 98.80Cq \hspace*{4.5cm} Preprint HUTP-96/A032, 
astro-ph/9608025}
\vskip2pc]

\section{Introduction}
\indent  There has been much activity lately in two exciting areas of
inflationary cosmology:  the construction of \lq\lq open inflation"
models, and the development of a new theory of post-inflationary
reheating.  To date,
these two developments have been treated separately; in this paper,
we explore some qualitatively new behavior for the reheating epoch in an
open universe. The open inflation
models \cite{Buch,YamOp,LinOp} exploit the fact, first noted
by Coleman and De Luccia and independently by Gott and
Statler~\cite{ColGott}, that the metric on the interior of a bubble
nucleated in a sea of de-Sitter-like false vacuum would be that of an {\it
open} Friedmann-Robertson-Walker spacetime.  By arranging for an epoch of
\lq\lq slow-roll" inflation to occur within the nucleated bubble, and
arranging for the number of {\it e}-folds of expansion during this second
inflationary phase not to exceed $N \sim 65$, it is possible to produce an
observable universe whose mass density today falls below the critical mass
density required for a flat universe:  $\Omega_o \sim 0.2 - 0.3$, rather
than the standard inflationary prediction, $\Omega_o = 1.0$. \\
\indent  Independently, several groups have re-examined recently the
process of particle production at the close of an inflationary phase, and
have found resonances and exponential instabilities in the decay of an
inflaton field into bosons, which had been overlooked in the original
accounts of reheating. \cite{KLS1,STB,Boyan,Yosh,DK96,Fuji,Reviews} 
According to the new theory of reheating, a post-inflationary second-order
phase transition should be viewed as having {\it three} distinct stages,
rather than the older two-stage picture:  the highly efficient, often
resonant, decay of the oscillating inflaton field into boson decay products
(which might include inflaton bosons themselves, due to the inflaton's
self-interaction), followed by the decay of these decay products (along the
methods originally developed in~\cite{oldreh}), followed finally by the
thermalization of these particles and the establishment of the
radiation-dominated era.  Kofman, Linde, and Starobinsky term this first,
resonant stage the \lq\lq preheating" stage.~\cite{KLS1} \\
\indent  As demonstrated in this paper, under certain conditions
the resonant decay of the inflaton in an open universe can
populate \lq\lq supercurvature" modes, which have no analogue in the flat
space case, and no particle-like
interpretation.  In some open inflation models the amplification of such
supercurvature modes can even dominate
the usual resonant creation of particles at
the time of reheating.  Thus, models of open inflation can display
dramatically different thermal histories following the end of inflation
from inflationary scenarios in a spatially flat universe.  \\
\indent  A Friedmann-Robertson-Walker universe with nonvanishing spatial
curvature has a physical curvature scale given by $a (t) / |K|$, where
$a(t)$ is the cosmic scale factor, and $K = -1$ or $+1$ for an open or
closed universe, respectively.  With $K = -1$, the comoving curvature scale
is thus simply $+1$.  From the Friedmann equation it can further be shown
that when $K = -1$ the physical curvature scale will always be larger 
than the Hubble distance, $H^{-1}$, for any cosmological epoch.
Eigenfunctions of the Laplacian operator can be found for this background
spacetime with eigenvalue $-(k/a)^2$, where $(k/a)$ is the inverse of a
physical length:  $k/a = 2\pi / \lambda_{ph}$, with $0 < k^2 < \infty$. 
Modes with comoving wavenumber $k^2 > 1$ thus vary on scales shorter than
the comoving curvature scale, and are labelled \lq\lq subcurvature" modes,
whereas modes with $0 < k^2 < 1$ correspond to \lq\lq supercurvature" modes. 
(See, {\it
e.g.},~\cite{LythW}.)  (Note that tensor modes can have different values
from scalar modes, such as $k^2 = -3$ for gravitational
waves.\cite{Garriga,GarcB,YST}) In a flat universe, the
comoving curvature scale runs off to infinity, so there are never
supercurvature modes.  It is this difference from the flat space case that
we examine here, for the epoch of preheating:  even though $\Omega
\rightarrow 1$ at the end of the second round of inflation, there will
still exist a comoving curvature length scale at $+1$, the existence of
which may change the usual physics at the time of preheating.  \\
\indent  We begin the study following the nucleation of a single bubble
in the midst of a sea of de Sitter space false vacuum.  Inside the
nucleated bubble, there is a nonvanishing scalar field potential and a
single relevant scalar field to drive a second round of inflation within
the bubble;  both the single-field models of open
inflation \cite{Buch,YamOp} and Linde's two-field \lq\lq hybrid"
open inflation model~\cite{LinOp} fit such a description.  
Inside the bubble, the Lagrangian density can thus be written:
%%%%%%%[equation 1]%%%%%
\beq
{\cal{L}} = \sqrt{-g} \left[ \frac{1}{16\pi G} R - \frac{1}{2} g^{\mu\nu} 
\partial_{\mu} \varphi \partial_{\nu} \varphi - V (\varphi) \right] ,
\label{lag}
\eeq
and the background spacetime takes the form~\cite{ColGott}
%%%%%%%%%[equation 2]%%%%
\beq
ds^2 = - dt^2 + a^2 (t) \left[ d\chi^2 + \sinh^2 \chi \left( d\theta^2 + 
\sin^2 
\theta d\phi^2 \right) \right] .
\label{ds}
\eeq
If we make the familiar decomposition,
%%%%%%%%[equation 3]%%%%%%
\beq
\varphi (t, \> \chi , \> \theta , \> \phi ) = \varphi_{o} (t) + 
\delta\phi (t, 
\> \chi , \>  \theta , \> \phi ),
\eeq
then the equation for the quantum fluctuations $\delta \phi$ becomes
%%%%%%%%%%[equation 4]%%%
\beq
\Box \left( \delta  \phi \right) - \left(\frac{\partial^2 V}{\partial 
\varphi^2}\right)_{\varphi_{o}} \left( \delta \phi \right)  = 0 .
\label{eom2}
\eeq
Switching to conformal time, $d\eta = a^{-1} dt$, and using the 
expression for 
the d'Alembertian, $\Box \varphi = (-g)^{-1/2} \partial_{\mu} \left[ 
(-g)^{1/2} 
g^{\mu\nu} \partial_{\nu} \varphi \right]$, then inside the bubble equation 
(\ref{eom2}) takes the form:
%%%%%%%%%[equation 5]%%%%
\beq
\left[ \frac{\partial^2}{\partial \eta^2} + 2 \frac{a^{\prime}}{a} 
\frac{\partial}{\partial \eta} - \vec{\rm{L}}^2 + a^2 \left( 
\frac{\partial^2 
V}{\partial \varphi^2} \right)_{\varphi_{o}} \right] \delta \phi = 0 ,
\label{eom3}
\eeq
where primes denote derivatives with respect to conformal time, $\eta$, 
and $\vec{\rm{L}}^2$ is the comoving Laplacian operator.  Eigenfunctions of
$\vec{\rm{L}}^2$ in the metric of equation (\ref{ds}) 
have been studied by many people (see, {\it
e.g.} \cite{YST,Lif}).  Denoting these harmonics as $Y_{klm}(\chi , \>
\theta, \> \phi )$, with $\vec{\rm{L}}^2 Y_{klm} = - k^2 Y_{klm}$, then we
may parametrize the fluctuations in terms of mode functions as: 
%%%%%%%[equation 6]%%%%%
\beq
\delta \phi_k (\eta , \> \chi , \> \theta , \> \phi ) = \frac{1}{a (\eta)} 
\psi 
(\eta) Y_{klm} (\chi , \> \theta , \> \phi ).
\eeq
(We will not need the explicit form of the $Y_{klm}$'s here, which can be
written in closed form in terms of associated Legendre
polynomials.~\cite{YST})  Finally, 
if we write $( \partial^2 V / \partial \varphi^2)_{\varphi_{o}}$ as 
${\cal{M}}^2 
(\varphi_{o})$, then the equation for the inflaton fluctuations becomes:
%%%%%%%[equation 7]%%%%%
\beq
\psi^{\prime\prime}_k - \frac{a^{\prime\prime}}{a} \psi_k + 
k^2 \psi_k + a^2 {\cal{M}}^2 \psi_k = 0 .
\label{eom4}
\eeq
(We have neglected the metric perturbations which couple to $\psi_k$,
which should be a good approximation during the reheating epoch; for more
on such metric perturbations, see~\cite{MFB}.)  In this paper, we will be
concerned with the behavior of solutions of equation (\ref{eom4}) after the
end of inflation, as the inflaton field $\varphi_o$ oscillates near the
minimum of its potential.\\
\indent  In Section II, we briefly review the conditions under which the
oscillating inflaton may decay resonantly into bosons.  In an open
universe, with the possibility of populating supercurvature modes during
the inflaton's parametric resonance, we must pay closer attention than is
usually done to the conditions under which the inflaton's decay may be
resonant:  the harmonic relations between the frequency of the inflaton's
oscillation and the decay products' momenta are specified here for all
resonance bands within the narrow resonance regime.  In Section III, these
resonance conditions are studied for the simplest model of an inflaton
decaying into inflaton bosons, due to a $\varphi^4$ self-coupling.  As will
be shown, in this case no supercurvature modes may be populated during
preheating, and there should be no deviation from the usual flat space
case.  The model is generalized in Section IV to the case of an inflaton
decaying into a distinct species of boson, via Yukawa and quartic
couplings, and in these cases supercurvature modes can be populated, if the
masses and couplings of the model satisfy certain relationships to very 
high accuracy.  None of the resonant behavior will affect inflaton decay 
into fermions, because of Fermi-Dirac statistics, and we will ignore such
couplings here.  The results are discussed in Section V.

\section{Preheating and Resonant Decays}
\indent  Near the minimum of its potential, the inflaton
will begin to oscillate, and this can set up a parametric resonance in the
inflaton's decay to bosons.~\cite{KLS1}
Assuming that the frequency of the inflaton's oscillations is greater than
the Hubble parameter at this time, 
we may neglect Hubble expansion during the course of
the resonant decays.  Then the inflaton field oscillates as $\varphi_o
(\eta) \simeq \overline{\varphi_o} \cos (\omega_{\rm osc} \eta )$, where
$\overline{\varphi_o}$ is slowly decreasing on the oscillation time-scale
$\omega_{\rm osc}^{-1} = (a_e m)^{-1}$.  Here $a_e$ is a constant giving the
value of the cosmic scale factor at the end of inflation.  Note that $(a_e
H_e )^{-1} < 1$, as will be true for any cosmological epoch in an open
universe.  Hence, our assumption about the frequency of the inflaton's
oscillation translates to $ (a_e m )^{-1} \ll ( a_e H_e )^{-1} < 1$, or
$a_e m \gg 1$.  The ${\cal{M}}^2
(\varphi_o )$ term then contains two types of terms, oscillating and
non-oscillating, and may be written:
%%%%%%%%%%%%%%%%[equation 8]%%%%%%%%%%%
\beq
{\cal{M}}^2 (\varphi_o ) = A + B \cos (q\omega_{\rm osc} \eta ) ,
\label{M2}
\eeq
where $A$, $B$, and $q$ are determined by a given model's particular
couplings.  The equation of motion for the inflaton fluctuations, equation
(\ref{eom4}), then becomes:
%%%%%%%%%%%%%%[equation 9]%%%%%%%%%%%%
\beqn
\nonumber  \psi_k^{\prime\prime} &+& \omega_k^2 \left[ 1 + g \cos
(q\omega_{\rm osc} \eta ) \right] \psi_k \simeq 0 , \\
\omega_k^2 &\equiv& k^2 + a_e^2 A \>\>,\>\> g \equiv a_e^2 B / \omega_k^2 .
\label{Mathieu}
\eeqn
This is now in the form of the Mathieu equation, solutions of which 
reveal exponential instabilities within various resonance 
bands.\footnote{As emphasized in \cite{Yosh,Fuji}, the exponential 
instabilities arise for
solutions of any differential equation with periodic coefficients, and are
more general than the particular form of equation 
(\ref{Mathieu}).} \\
\indent  The narrow resonance
regime of equation (\ref{Mathieu}), 
with $g < 1$, can be studied perturbatively.  The analytic
analysis in \cite{STB,Yosh,DK96} demonstrates 
that in this case the
solutions are exponentially unstable in narrow resonance bands $n
q \omega_{\rm osc}  =
2\omega_k \pm \frac{1}{2} |\epsilon_n |$, with $|\epsilon_n| < \omega_k$,
and $n$ a positive integer.  Each resonance band with $n > 1$ corresponds
to keeping terms of $O (g^n)$ in the solution to equation (\ref{Mathieu}),
and the width of each successive resonance band in the narrow resonance
regime shrinks as~\cite{L2}:
%%%%%%%%%%%%[equation 10]%%%%%%%%%%%
\beq
|\epsilon_n | = \frac{n^{2n - 3} g^n \omega_k}{2^{3 (n - 1)} \left[ (n -
1)! \right]} \equiv b_n g^n \omega_k .
\label{epsilon_n}
\eeq
Near the center of each resonance band, solutions take the form
%%%%%%%%%%%%[equation 11]%%%%%%%%%%
\beq
\psi_{k , \> n}^{\pm} (\eta) \simeq \frac{ \exp (\pm s_k^{(n)} \eta
)}{\sqrt{ 2 \omega_k }} ,
\label{psi_k}
\eeq
with $s_k^{(n)}$ real and positive.
Modes with a given $k^2$ slide out of the resonance band due to the slow
decrease of $\overline{\varphi_o}$, and hence of $g$, due to scatterings and
back-reaction effects.  Given such a time-dependence of $g$, modes will
only become resonantly amplified if they initially satisfy 
%%%%%%%%%%%%[equation 12]%%%%%%%%%
\beq
n q \omega_{\rm osc} = 2 \omega_k + \frac{1}{2} |\epsilon_n | ;
\label{resband}
\eeq
modes initially satisfying $n q\omega_{\rm osc} = 2 \omega_k - \frac{1}{2}
|\epsilon_n |$, on the bottom edge 
of the resonance band, will not remain within
the band long enough to become amplified.  We will thus study equation
(\ref{resband}) for various models to determine which areas of parameter
space allow for the production and amplification of supercurvature modes. 
The subcurvature modes should behave as in the flat space case, and hence
we will not treat them further here. \\
\indent  As demonstrated explicitly
in~\cite{DK96}, including a nonzero Hubble expansion during the period over
which the inflaton oscillates does not change the qualitative form of these
solutions, but merely narrows the width of the resonance bands.  Similarly,
including back-reaction and rescattering effects should further decrease
the width of each resonance
band. \cite{KLS1,Boyan,Yosh,Fuji,Khleb,Campbell}  We may accommodate these 
effects by inserting a phenomenological
parameter $\alpha (\eta )$, with $0 < \alpha (\eta) < 1$, into the band
width, writing the width of each resonance band as $|\epsilon_n | = b_n
\alpha^n (\eta ) g^n \omega_k$, though we will set $\alpha = 1$ in the
following.  \\
\indent  The broad resonance regime of equation (\ref{Mathieu}) ($g \gg
1$), identified in~\cite{KLS1}, has been studied analytically
in~\cite{Fuji}, and also leads to solutions of the form 
$\psi_{k, \> n}^{\pm} \propto \exp (\pm \sigma_k^{(n)} \eta )$, with
$\sigma_k^{(n)}$ real and positive (and $\sigma_k^{(n)} \gg s_k^{(n)}$). 
However, such a broad resonance regime is ruled out for most quartic
couplings \cite{Fuji}; and even for models with Yukawa couplings
between the inflaton and a distinct species of boson, the coupling strength
(and hence $g$) may not be made arbitrarily large, because such large
couplings would interrupt the usual inflationary dynamics prior to the
preheating epoch.  Furthermore, the particles produced during a broad
resonance have much higher momenta than particles produced in the narrow
resonance regime, and hence no supercurvature modes are expected to be
populated from a broad resonance.  For these reasons, we will restrict
attention here to the narrow resonance regime for each of the models
studied below.

\section{Inflaton Decay Into Inflaton Bosons}
\indent  In this section, we consider a model in which the oscillating
inflaton field decays into inflaton bosons, due to a $\varphi^4$
self-coupling.  Near the minimum of its potential, any model of open
inflation should assume a general chaotic inflation form~\cite{chaotic}:
%%%%%%%%%%%%[equation 13]%%%%%%%%%
\beq
V (\varphi ) = \frac{1}{2} m^2 \varphi^2 + \frac{1}{4} \lambda \varphi^4 .
\label{Vphi4}
\eeq
In this case, the parameters in equation (\ref{Mathieu}) become
%%%%%%%%%%%%%%%[equation 14]%%%%%%%%%
\beq
\omega_k^2 = k^2 + a_e^2 m^2 + \frac{3}{2} a_e^2 \lambda 
\overline{\varphi_o}^2
\>\> , \>\> g = \frac{3}{2} a_e^2 \lambda \overline{\varphi_o}^2 /
\omega_k^2 ,
\label{omegakgphi4}
\eeq
or, combining terms,
%%%%%%%%%%%%[equation 15]%%%%%%
\beq
\omega_k^2 = \frac{\left( k^2 + a_e^2 m^2 \right)}{(1 - g)} .
\label{omegak2}
\eeq
Note from the form of $g$ in equation (\ref{omegakgphi4}) that $g < 1$
always in this model:  in the case of inflaton decay into inflaton bosons,
the only possible resonances lie in the narrow resonance regime, and hence
the explicit expressions in section II may be used. 
Finally, for the potential of equation (\ref{Vphi4}), the 
oscillating term in
equation (\ref{Mathieu}) oscillates at frequency $2 \omega_{\rm osc} = 2
a_e m$, or $q = 2$, after use has been made of a double-angle formula for
$\cos^2 (\omega_{\rm osc} \eta )$.  \\
\indent  Given that $a_e m \gg 1$, we need only consider wavenumbers in the
range $0 < k \ll a_e m$ in our search for any supercurvature modes which may
be resonantly amplified during preheating.  Defining
%%%%%%%%%%%%[equation 16]%%%%%%%%%%
\beq
\ell \equiv \frac{k}{a_e m} ,
\label{ell}
\eeq
then equation (\ref{resband}) becomes:
%%%%%%%%%%%%%%[equation 17]%%%%%%%%
\beqn
\nonumber 2 \left( n\sqrt{1 - g} - 1 - \frac{1}{4} b_n g^n \right) = \\
\left[ \ell^2 - \frac{1}{4} \ell^4 + O (\ell^6 ) \right] \left(1 +
\frac{1}{4} b_n g^n \right) .
\label{n1}
\eeqn
Keeping only the lowest-order terms in $\ell$, this gives 
%%%%%%%%%%%%%%%%[equation 18]%%%%%%
\beq
k_{\rm res}^2 \simeq 2 a_e^2 m^2 \left[ \frac{4n \sqrt{1 - g} - 4 - b_n
g^n}{4 + b_n g^n} \right] .
\label{kres1}
\eeq
When $n = 1$, the quantity on the righthand side of equation (\ref{kres1})
is negative, revealing that neither supercurvature nor subcurvature modes
will be amplified in the lowest resonance band when $k < a_e m$. 
For $n > 1$, the approximation $\ell \ll 1$ is 
violated, and thus equation (\ref{kres1})
cannot be used to study resonant subcurvature modes. 
Instead, it can easily be demonstrated that when $\ell > 1$ 
(and hence for subcurvature modes only), resonant modes correspond to
%%%%%%%%%%%%%%[equation 19]%%%%%%%
\beqn
\nonumber k_{\rm res} &\simeq& \frac{n \sqrt{1 - g}}{2\gamma} a_e m \left[ 1
\pm  \sqrt{1 - \frac{2 \gamma^2}{n^2 (1 - g)}} \right] + O(k
\ell^{-2}), \\
\gamma &\equiv& 1 + \frac{1}{4} b_n g^n \geq 1 ,
\label{kres2}
\eeqn
again revealing that no consistent solution (with $k_{\rm res}$ real and
positive) may be found for $n = 1$.  Thus, for the potential of equation
(\ref{Vphi4}), only subcurvature modes may be amplified during preheating,
and only in the narrow resonance regime with $n \geq 2$.  \\
\indent  For this particular potential, a limit on $g$ may be set by the
independent constraints on $m$ and $\lambda$
from observed microwave background anisotropies:  $\lambda \leq 10^{-14}$,
and $m / M_{pl} \leq 10^{-6}$.~\cite{MFB}  An upper bound may be placed on
$\overline{\varphi_o}$ by equating this with the value of $\varphi_o$ once
inflation ends, that is, once the slow-roll conditions are violated, at 
$\partial^2 V / \partial \varphi^2 \simeq 24 \pi M_{pl}^{-2} V$.  This gives
$\overline{\varphi_o} \leq 0.16 M_{pl}$ (using the constraints on $\lambda$
and $m$), or $\lambda \overline{\varphi_o}^2 / m^2 \leq 2.6 \times 10^{-4}$,
guaranteeing that $g \leq 4 \times 10^{-4}$.  With so small a $g$,
resonance bands with $n > 2$ are unlikely to produce large occupation
numbers of inflaton bosons, and the resonant decay of the inflaton may be
rather inefficient in transferring the energy density of the decaying
inflaton field into boson decay products.  Such tight constraints on the
potential's parameters are absent in the case of inflaton decay into a
distinct species of boson, assuming that the second boson field
is unimportant
during the inflationary phase (when the primordial density perturbation
spectrum is determined).  It is to these models that we now turn.

\section{Inflaton Decay Into A Distinct Boson Species}
\indent  In the case of inflaton decays into a distinct species of boson,
$\chi$, we may study a potential of the form:
%%%%%%%%%%%%%%%%%[equation 20]%%%%%%
\beq
V (\varphi , \> \chi ) = \frac{1}{2} m^2 \varphi^2 + \frac{1}{2} m_{\chi}^2
\chi^2 + \mu \varphi \chi^2 + \lambda_{\chi}^2 \varphi^2 \chi^2 .
\label{V2}
\eeq
In supersymmetric models, these couplings will satisfy $\mu = 2
\lambda_{\chi} m$.~\cite{Campbell}
Writing $\chi = a^{-1} (\eta ) f (\eta ) Y_{plm} (\vec{\rm x})$, the equation
for $f (\eta )$ will assume the same form as equation (\ref{eom4}), with
${\cal{M}}^2 \rightarrow \partial^2 V / \partial \chi^2 = m_{\chi}^2 + 2
\mu \varphi + 2 \lambda_{\chi}^2 \varphi^2$.  Again taking $\omega_{\rm
osc}^{-1} = (a_e m)^{-1} \ll (a_e H)^{-1}$, the $\chi$-modes will obey:
%%%%%%%%%%%%%[equation 21]%%%%%%%%
\beqn
\nonumber f_k^{\prime\prime} &+& \omega_{\chi}^2 
\left[ 1 + h_1 \cos (2\omega_{\rm osc}
\eta ) + h_2 \cos (\omega_{\rm osc} \eta ) \right] f_k \simeq 0 , \\
\nonumber \omega_{\chi}^2 &\equiv& k^2 + a_e^2 m_{\chi}^2 + 
a_e^2 \lambda_{\chi}^2 \overline{\varphi_o}^2 , \\
h_1 &\equiv&  a_e^2 \lambda_{\chi}^2 \overline{\varphi_o}^2 /
 \omega_{\chi}^2 \>\>,\>\>
h_2 \equiv 2 a_e^2 \mu \overline{\varphi_o}/\omega_{\chi}^2 .
\label{Mathieu2}
\eeqn
Note from the form of $h_1$ that $h_1 < 1$ always, so only the narrow
resonance regime is viable for the quartic coupling.  If we
first consider $h_2 \rightarrow 0$, and if the mass difference between the
inflaton and the $\chi$ boson is small, then modes with $k \ll a_e
m_{\chi}$ will become exponentially amplified in bands centered on:
%%%%%%%%%%%[equation 22]%%%%%%
\beq
k_{\rm res}^2 \simeq 2 a_e^2 m_{\chi}^2 \left[ \frac{4n 
( m / m_{\chi} ) \sqrt{1 - h_1} - 4 - b_n h_1^n }{4 + b_n h_1^n}
\right] .
\label{kres3}
\eeq
The next order term is of $O(k^4 a_e^{-2} m_{\chi}^{-2})$, which will be
completely negligible if $0 < k^2 < 1$ and $a_e m_{\chi} \gg 1$.
Though the form of equation (\ref{kres3})
for $k_{\rm res}^2$ looks quite similar to the
corresponding expression for $\varphi \rightarrow \varphi$ decays, equation
(\ref{kres1}), there are important differences.  First, there are now two
free parameters, $(m / m_{\chi})$ and $h_1$, rather than only the single
free parameter $g$ in equation (\ref{kres1}).  Second, the coupling
constant $\lambda_{\chi}$ (and hence $h_1$) is not limited directly by
today's observed microwave background spectrum, because it is assumed that
the $\chi$-field plays no role during inflation, so $h_1$ need not be
constrained to $h_1 \sim O (10^{-4})$ as $g$ is.  Finally, the ratio $(m /
m_{\chi})$ may be either greater or less than unity:  as noted
in~\cite{KLS1}, it is possible for the inflaton to decay into a particle
heavier than itself during the resonant preheating stage (a decay which is
kinematically forbidden according to non-resonant Born theory).  Thus,
supercurvature modes will be produced and resonantly amplified in a
resonance band $n_*$ which satisfies
%%%%%%%%%%%%%%[equation 23]%%%%%
\beq
0 < \left[ \frac{8n_* (m / m_{\chi}) \sqrt{1 - h_1} - 8 - 2 b_{n_*}
h_1^{n_*}}{4 + b_{n_*} h_1^{n_*}} \right] < \left( a_e m_{\chi}
\right)^{-2} .
\label{n*1}
\eeq
If $a_e \sim e^{65} \sim 10^{28}$, then this is a very tiny window: the
resonant production of supercurvature modes requires that the masses and
couplings of the potential obey certain interrelationships to exponential
accuracy.  Still, by exploiting the freedom in $(m / m_{\chi})$ and
$\lambda_{\chi}$, it is possible to arrange for such conditions to be met
(and the nearly-exact cancellation of combinations of parameters is not in
itself foreign to particle physics).  Furthermore, if $n_* = 1$ and $h_1
\ll 1$, then all higher (subcurvature) modes could fall outside of
resonance bands, leaving only the supercurvature modes to be produced and
amplified at the preheating stage.  Note that if $m_{\chi} \ll m$, then
equation (\ref{kres3}) is inappropriate, and only subcurvature modes will
be amplified at preheating.  Other resonant subcurvature modes, with $k >
a_e m_{\chi}$, may be found analogously to equation (\ref{kres2}), and will
not be pursued here. \\
\indent  Similarly, if $h_1 \rightarrow 0$ and the Yukawa coupling in
equation (\ref{V2}) is active ($h_2 \neq 0$), then supercurvature modes
will be produced resonantly in a resonance band $n_*$ if
%%%%%%%%%%%%%%%%%%[equation 24]%%%%%%%%
\beq
0 < \left[ \frac{4 n_* (m / m_{\chi}) - 8 - 2 b_{n_*} h_2^{n_*} }{4 +
b_{n_*} h_2^{n_*} } \right] < \left( a_e m_{\chi} \right)^{-2}
\label{kres4}
\eeq
in the narrow resonance regime, up to terms of $O(k^4 a_e^{-2}
m_{\chi}^{-2})$.  Again, the pair of parameters $(m /
m_{\chi})$ and $\mu$ may be tuned to satisfy this condition, and if $n_* =
1$ and $h_2 \ll 1$, then the production and amplification of supercurvature
modes will dominate the preheating phase.  As for the $h_1$ case, if 
$m_{\chi} \ll m$ or if $k > a_e m_{\chi}$, then only subcurvature modes
will be produced at preheating. \\
\indent  Care must thus be taken to identify the (model-dependent) 
dominant decay
channel of the inflaton, especially if the two couplings in equation
(\ref{V2}) are in competition with the inflaton self-coupling in equation
(\ref{Vphi4}).  The conditions under which the supercurvature modes may
become amplified and dominate the production of subcurvature modes 
during the preheating phase depends sensitively upon the couplings and
mass scales of the decaying inflaton and its decay-product bosons. Note
that although $k_{\rm res}^2 \rightarrow 0$ for these supercurvature modes,
there is no infrared catastrophe, since every resonance band in the narrow
resonance regime has a finite band width.

\section{Discussion}
\indent  What do these exponentially amplified supercurvature modes
represent?  In the usual flat space preheating scenario, narrow resonance
modes simply correspond to the production of low-momentum (small $k$)
particles, whose occupation numbers may be estimated with the aid of a
Bogolyubov transformation.  Yet in the open universe, such small-$k$ modes
can stretch beyond the curvature scale (and hence beyond the horizon), and no
particle interpretation may be given to them.  (This is a specific example
of how tenuous the notion of \lq\lq particle" can become in general curved
spacetimes.~\cite{qft})\footnote{As pointed out in~\cite{LythW},
supercurvature modes should be absent from the mode expansion of
any quantum field in its vacuum
state, since any square-integrable function may be defined completely in
terms of subcurvature modes; the production of supercurvature
modes at preheating is consistent
with the inflaton's departure from the pure vacuum state at that time.}
Instead, they will contribute to (very) long-wavelength curvature
perturbations, akin to supercurvature modes produced during
inflation \cite{Garriga,GarcB,YST}, though their evolution
will be complicated by the highly non-adiabatic nature of the reheating
epoch.  The detailed evolution of these modes may be tracked using the
methods of~\cite{MFB}, though it is unlikely that they will contribute
greatly to any observed microwave background anisotropies today, 
given their very long
correlation lengths~\cite{LythW}; this is a subject of further study. \\
\indent  Rather, their most dramatic role could be in changing the thermal
history immediately following reheating:  the new reheating scenario
depends upon the explosive production of particles, far from thermal
equilibrium; these decay products then thermalize via interactions on a
(potentially) quite different time-scale.  When the occupation numbers of
out-of-equilibrium decay-product particles becomes quite large, this opens 
up the possibility
for such microphysical processes as non-thermal phase 
transitions~\cite{NonT}, preheating-induced supersymmetry
breaking~\cite{susy}, and GUT-scale baryogenesis~\cite{GUT}.  Yet the 
foregoing
analysis reveals that in an open universe, the (resonant) preheating
production of particles may be either totally absent or strongly 
diluted, due to some
portion of the inflaton's energy density being \lq siphoned off' into
supercurvature modes.  For models in which the resonant production of
subcurvature modes is subdominant, the inflaton will 
still decay into
particles and populate the universe following the end of the resonance
regime, according to the processes of the older theory of reheating.  
If this resonant production of supercurvature modes and non-resonant
production of particles becomes the dominant effect, then these
other microphysical processes in the post-preheating epoch may need to be
reconsidered. \\
\indent  Given the exponentially tiny supercurvature resonance regimes in
equations (\ref{kres3}) and (\ref{kres4}), however, it is rather unlikely
(though still possible) that the supercurvature modes from the resonant 
$\varphi \rightarrow \chi$ decays would dominate the preheating epoch.  And
such modes are completely absent from a strictly $\varphi \rightarrow
\varphi$ preheating epoch.  Thus, large deviations from preheating
scenarios in flat space are unlikely to appear for models of open
inflation. \\
\indent  Finally, as noted in the conclusion in~\cite{DK96}, for models
with non-minimally coupled scalar fields which recover the Einstein-Hilbert
gravitational action near the end of inflation ({\it e.g.}, models of the
form studied in~\cite{DK95}), the analysis of reheating carries over 
unchanged. Thus, the recent open inflation model based on induced-gravity
theory~\cite{igiop} should display the same resonance features and
supercurvature modes as analyzed above. 

\section*{Acknowledgments}
\indent  It is a pleasure to thank Alan Guth and Jaume Garriga for helpful
discussions during the course of this work.  This research was supported by
the NSF.

%_________________________________

%

\end{document}